\documentclass[10pt,aps,prd,superscriptaddress,longbibliography]{revtex4}   %eqsecnum,
\usepackage{mathrsfs,amsmath,amsthm,latexsym,amssymb,amsfonts,epsfig,txfonts,cancel}
\newcommand{\makeSymbol}[1]{\mathord{\vcenter{\hbox{#1}}}}
\usepackage{xcolor}
\definecolor{navy}{rgb}{0.0,0.06,0.5}
\usepackage[colorlinks, linkcolor=cyan, citecolor=magenta, urlcolor=navy, plainpages=false, pdfstartview=FitH]{hyperref}
\usepackage{appendix}
\allowdisplaybreaks
\newcommand{\GZU}{School of Physics, Guizhou University, Guiyang 550025, China}
\newcommand{\BNU}{Department of Physics, Beijing Normal University, Beijing 100875, China}
\newcommand{\sst}{\scriptscriptstyle}
\begin{document}

\title{Consistency check on the fundamental and alternative flux operators in loop quantum gravity}

%-------------------------------------------------------------
\author{Jinsong Yang}
\email{jsyang@gzu.edu.cn}
\affiliation{\GZU}

\author{Yongge Ma}
\thanks{Corresponding author}
\email{mayg@bnu.edu.cn}
\affiliation{\BNU}
%-------------------------------------------------------------

\begin{abstract}
There are different constructions of the flux of triad in loop quantum gravity, namely the fundamental and alternative flux operators. In parallel to the consistency check on the two versions of operator by the algebraic calculus in the literature, we check their consistency by the graphical calculus. Our calculation based on the original Brink graphical method is obviously simpler than the algebraic calculation. It turns out that our consistency check fixes the regulating factor $\kappa_{\rm reg}$ of the Ashtekar-Lewandowski volume operator as $\frac12$, which corrects its previous value in the literature.
\end{abstract}
%------------------------------------------------------------- 
\maketitle
%------------------------------------------------------------- 

\section{Introduction}

Loop quantum gravity (LQG) takes the key lesson from general relativity (GR) that the spacetime geometry is dynamic rather than static to build a background independent quantum theory of gravity, which has made an outstanding impact in the field (see \cite{Rovelli:2004tv,Thiemann:2007pyv} for books, and \cite{Thiemann:2002nj,Ashtekar:2004eh,Han:2005km,Giesel:2012ws} for articles). Two formulations, the canonical (Hamiltonian) and covariant (Lagrangian) formulations are being studied in LQG. In the canonical formulation, the kinematical representation of the holonomy-flux algebra is shown to be unique to certain sense \cite{Lewandowski:2005jk}, and the geometric operators corresponding to length, area, and volume functions are constructed and all have discrete spectrum \cite{Rovelli:1994ge,Ashtekar:1996eg,Ashtekar:1997fb,Thiemann:1996at,Ma:2010fy}. An open problem of LQG is how to implement the quantum dynamics. Approaching to this problem in the canonical formulation, some mathematically well-defined Hamiltonian constraint operators were constructed to determine quantum dynamics \cite{Thiemann:1996aw,Yang:2015zda,Alesci:2015wla}, and their key properties were also studied \cite{Alesci:2011ia,Thiemann:2013lka,Zhang:2018wbc,Zhang:2019dgi}. Moreover, the non-perturbative quantization technique was also extended to define the Hamiltonian constraint operators for other important alternative theories of gravity \cite{Zhang:2011vi,Zhang:2011qq,Zhang:2011vg,Zhang:2011gn,Ma:2011aa}. In the covariant formulation, some reasonable transition amplitudes were also proposed \cite{Engle:2007qf,Freidel:2007py,Kaminski:2009fm,Ding:2010fw}.

The first mathematically well-defined Hamiltonian constraint operator for pure gravity was constructed in the canonical LQG by Thiemann using the cotriad operator \cite{Thiemann:1996aw}, which is often called Thiemann's trick in the literature. Moreover, the cotriad operator was also applied to construct densely defined Hamiltonian constraint operators for gravity coupled to matters \cite{Thiemann:1997rt}, as well as a length operator \cite{Thiemann:1996at}. In order to enhance the confidence in employing the cotriad operator to construct the Hamiltonian constraint, a consistency check was proposed at the kinematical level by comparing the action of the alternative flux operator defined by the cotriad operator with the one of the fundamental flux operator on the same state \cite{Giesel:2005bk,Giesel:2005bm}. Furthermore, similar ideals were recently adopted to define new alternative volume and inverse volume operators for LQG by using the cotriad operator \cite{Yang:2016kia}. Both the volume and inverse volume operators in \cite{Yang:2016kia} share the same qualitative properties with the volume operator defined in \cite{Ashtekar:1997fb}. To implement these consistency checks, one need to compute in detail the actions of these operators on the quantum states. Obviously, it is important to choose a suitable method for the calculus.

Recently, the graphical calculus based on the original Brink graphical method has been systematically applied to LQG \cite{Brink:1968bk,Yutsis:1962bk,Yang:2015wka}. The graphical method provides a very powerful technique for simplifying the complicated calculations. In this paper, the graphical calculus will be used to check on the consistency between the alternative flux operator and the fundamental flux operator, which was also studied by the algebraic calculus in \cite{Giesel:2005bk,Giesel:2005bm}. Comparing to the algebraic method, our derivation is obviously more compact and simple. Moreover, our result corrects the value of the regulating factor $\kappa_{\rm reg}$ of the volume operator in the literature.

\section{Consistency check on the fundamental and alternative flux operators}
In this section, we briefly summarize the elements of LQG. Then we introduce the construction of the fundamental and alternative flux operators. The consistency check on them will be studied in detail by employing the graphical calculus.

\subsection{The fundamental and heuristic alternative flux operators}
In the canonical LQG, the 4-dimensional spacetime manifold $M$ is split into $M=\varmathbb{R}\times \Sigma$ with $\Sigma$ being a 3-dimensional manifold of arbitrary topology. GR can be casted in the Hamiltonian formulism as a dynamical theory of the Ashtekar-Barbero connection with $SU(2)$ gauge group. The canonical variables are the $SU(2)$ connection $A^i_a:=\Gamma^i_a+\beta K^i_a$ and the densitized triad $\tilde{E}^a_i:=\sqrt{\det{q}}\,e^a_i$ on $\Sigma$, where spatial indices are denoted by $a,b,c,\cdots$ and $i,j,k,\cdots=1,2,3$ are internal indices, $\Gamma^i_a$ is the spin connection on $\Sigma$, $\beta$ is the Barbero-Immirzi parameter, $K^i_a$ is the extrinsic curvature of $\Sigma$, $\det(q)$ denotes the determinant of the three-metric $q_{ab}$ on $\Sigma$, and $e^a_i$ is the triad. The only nontrivial Poisson bracket between these canonical variables reads
\begin{align}\label{eq}
\{A^i_a(x),\tilde{E}^b_j(y)\}=\kappa\,\beta\,\delta^b_a\delta^i_j\delta^3(x,y)\,,
\end{align}
where $\kappa=8\pi G$ with $G$ being the usual gravitational constant. The fundamental variables for LQG are the holonomy $h_e(A)$ of $A^i_a$ along an 1-dimensional curve (edge) $e: [0,1]\rightarrow\Sigma$ and the flux $\tilde{E}_i(S)$ of $\tilde{E}^a_i$ through a 2-dimensional surface $S$. It is shown that the diffeomorphism invariant representation, the Ashtekar-Isham-Lewandowski representation, of holonomy-flux is unique to certain sense \cite{Lewandowski:2005jk}. The unique representation space, called also the kinematical Hilbert space, is ${\cal H}_{\rm kin}=L^2(\bar{\cal A},{\rm d}\mu_o)$, where $\bar{\cal A}$ is the space of distributional connections, and ${\rm d}\mu_o$ is the Ashtekar-Lewandowski measure \cite{Ashtekar:1991kc,Ashtekar:1994mh}. The typical elements of ${\cal H}_{\rm kin}$ is the so-called cylindrical functions $f_\gamma$ of $A\in \bar{\cal A}$ with respect to a graph $\gamma$. The spin network states provide the basis of ${\cal H}_{\rm kin}$ \cite{Rovelli:2004tv,Thiemann:2007pyv,Thiemann:2002nj,Ashtekar:2004eh,Han:2005km,Giesel:2012ws}.

A holonomy function is directly quantized as a  multiplication operator on ${\cal H}_{\rm kin}$. The flux $\tilde{E}_j(S)$ through a surface $S$ can also be quantized as the fundamental flux operator $\hat{\tilde{E}}^{\rm Fun}_i(S)$ by first implementing suitable regularization and then replacing $\tilde{E}^a_i$ by its quantum distribution $\hat{\tilde{E}}^a_i:=-{\rm i}\hbar\kappa\beta\delta/\delta A^i_a$ \cite{Ashtekar:1996eg,Thiemann:2007pyv}. Given a graph $\gamma$ and a surface $S$ on $\Sigma$, by changing the orientations of some edges of $\gamma$ and splitting edges of $\gamma$ into two halves at an interior point if necessary, we can obtain a graph $\gamma_{ S}$ adapted to $S$ such that the edges of $\gamma_{ S}$ belong to the four types: (i) $e$ is up w.r.t $S$ if $n_a^S(e(0))\dot{e}^a(0)>0$; (ii)  $e$ is down w.r.t $S$ if $n_a^S(e(0))\dot{e}^a(0)<0$; (iii) $e$ is inside w.r.t $S$ if $e\cap S=e$; (iv) $e$ is outside w.r.t. $S$ if $e\cap S=\emptyset$. Here $n^S_a$ is the co-normal with respect to $S$, and $\dot{e}^a(t)$ denotes the tangent vector of $e$. Then the flux operator $\hat{\tilde{E}}^{\rm Fun}_i(S)$ acting on a function $f_\gamma$ cylindrical with respect to a graph $\gamma$ adapted to $S$ is given by \cite{Ashtekar:1996eg,Rovelli:2004tv,Thiemann:2007pyv,Ashtekar:2004eh,Han:2005km}
\begin{align}\label{fun-action-i}
 \hat{\tilde{E}}^{\rm Fun}_i(S)\,\cdot f_\gamma&=\frac{\ell_{\rm p}^2\,\beta}{2}\sum_{v\in \gamma\cap S}\sum_{b(e)=v}\varrho(e,S)J^i_e\,\cdot f_\gamma\,,
\end{align}
where $\ell_{\rm p}^2\equiv \hbar\kappa$, the factor $\varrho(e,S)$ takes the values of $0$, $+1$ and $-1$ corresponding to the edge $e$ is inside/outside, up or down with respect to the surface $S$, the first sum is over the intersecting points $v$ between $\gamma$ and $S$, and the second sum is over those edges which have $v$ as a beginning point, and $J^i_e$ is the self-adjoint operator of the right-invariant vector field on the copy of $SU(2)$ corresponding to the edge $e$.

\begin{figure}[t]
\centering
\begin{minipage}[b]{0.2\textwidth}
\begin{tabular}{c}
\includegraphics[]{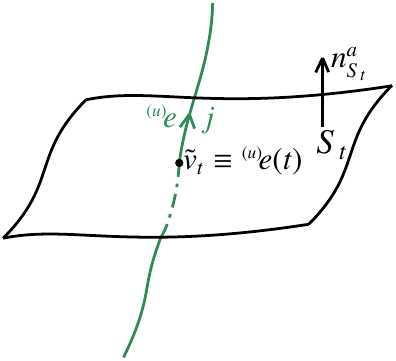}\\
(a)
\end{tabular}
\end{minipage}
\hspace{0.3cm}
\begin{minipage}[b]{0.2\textwidth}
\begin{tabular}{c}
\includegraphics[]{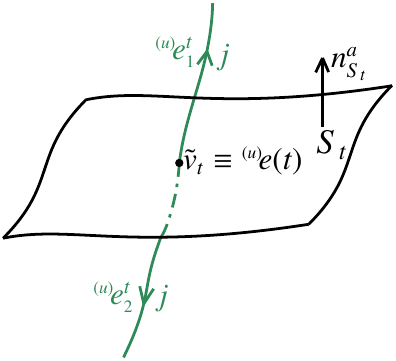}\\
(b)
\end{tabular}
\end{minipage}
\caption{(a) An edge ${}^{\sst (u)}\!e$ intersects a surface $S_t$ at $\tilde{v}_t\equiv {}^{\sst (u)}\!e(t)$. (b) The original edge ${}^{\sst (u)}\!e$ is partitioned by $S_t$ into two edges ${}^{\sst (u)}\!e^t_1$ and ${}^{\sst (u)}\!e^t_2$ starting from $\tilde{v}_t$.}
\label{fig-up}
\end{figure}

Alternatively, the classical flux function can also be expressed in terms of the cotriad $e^i_b$, since the densitized triad is related to the cotriad by $\tilde{E}^a_i=\frac12\epsilon_{ijk}\tilde{\epsilon}^{abc}{\cal S}e^j_be^k_c$, where $\tilde{\epsilon}^{abc}$ is the Levi-Civita tensor tensity of weight $1$ and ${\cal S}\equiv{\rm sgn}[\det(e^i_a)]$. Therefore, it can be quantized as an alternative flux operator using the cotriad operator \cite{Giesel:2005bk,Giesel:2005bm}. We now introduce the construction of the heuristic alternative flux operator. Consider an edge, in a graph $\gamma$, ${}^{\sst (u)}\!e:[0,1]\rightarrow \Sigma$, isolated intersects a surface denoted by $S_t$ at an additional vertex $\tilde{v}_t\equiv {}^{\sst (u)}\!e(t)$ with ${}^{(u)}\!\dot{e}^a(t)n_a^{S_t}>0$, and it is subdivided into two edges ${}^{\sst (u)}\!e^t_1$ and ${}^{\sst (u)}\!e^t_2$ starting from $\tilde{v}_t$ (see Fig. \ref{fig-up}). Then the classical flux can be expressed in terms of the cotriad $e^i_b$ as \cite{Giesel:2005bk,Giesel:2005bm}

\begin{figure}[t]
\centering
\includegraphics[height=0.25\textwidth]{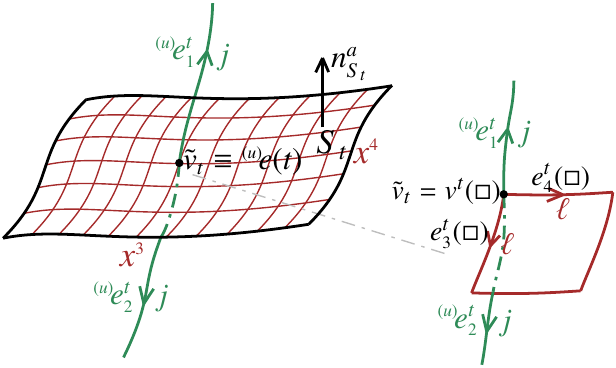}\\
\caption{A partition ${\cal P}_{\epsilon'} (S_t)$ of $S_t$ adapted to the coordinates $\{x^3,x^4\}$ into boxes $\Box$ with area $\epsilon'^2$, in which each box $\Box$ contains a vertex $v^t(\Box)$ and the edges $e^t_3(\Box)$ and $e^t_4(\Box)$ starting from $v^t(\Box)$ along the $x^3$ and $x^4$ coordinate lines, respectively, with length $\epsilon'$. Moreover, the partition ${\cal P}_{\epsilon'} (S_t)$ is also required to adapted to the graph $\gamma$ in a way that the graph $\gamma$ intersects at least one $\Box\in\,{\cal P}_{\epsilon'}(S_t)$ at $\tilde{v}_t=v^t(\Box)$.}
\label{partition}
\end{figure}

\begin{widetext}
\begin{align}\label{class-alt-reg-i}
\tilde{E}^{\rm Alt}_i(S_t)&=\int_{S_t}(*\tilde{E}_i)_{bc}=\int_{S_t}{\cal S}\epsilon_{ijk}e^j_be^k_c
=\int_{S_t}{\rm d}x^3{\rm d}x^4{\cal S}\epsilon_{ijk}e^j_3 e^k_4=\left(\frac{2}{\kappa\beta}\right)^2\int_{S_t}{\rm d}x^3{\rm d}x^4{\cal S}\epsilon_{ijk}\left\{A^j_3,V\right\}\left\{A^k_4,V\right\}\notag\\
&=\left(\frac{2}{\kappa\beta}\right)^2\lim_{\epsilon'\rightarrow0}\sum_{\Box\in\,{\cal P}_{\epsilon'}(S_t)}\int_\Box{\rm d}x^3{\rm d}x^4{\cal S}\epsilon_{ijk}\left\{A^j_3,V_{v^t(\Box)}\right\}\left\{A^k_4,V_{v^t(\Box)}\right\}\notag\\
&=24\left[\kappa\,\beta\,\chi(\ell)\right]^{-2}\lim_{\epsilon'\rightarrow0}\sum_{\Box\in\,{\cal P}_{\epsilon'}(S_t)}\int_\Box{\rm d}x^3{\rm d}x^4{\rm Tr}_\ell\left(\left\{A^j_3\tau_j,V_{v^t(\Box)}\right\}\tau_i{\cal S}\left\{A^k_4\tau_k,V_{v^t(\Box)}\right\}\right)\notag\\
&=24\left[\kappa\,\beta\,\chi(\ell)\right]^{-2}\lim_{\epsilon'\rightarrow0}\sum_{\Box\in\,{\cal P}_{\epsilon'}(S_t)}{\rm Tr}_\ell\left(h_{e^t_3(\Box)}\left\{h_{e^t_3(\Box)}^{-1},V_{v^t(\Box)}\right\}\tau_i\,{\cal S}\, h_{e^t_4(\Box)}\left\{h_{e^t_4(\Box)}^{-1},V_{v^t(\Box)}\right\}\right)\,,
\end{align}
\end{widetext}
where $*$ denotes the Hodge dual, $\chi(\ell)\equiv\sqrt{\ell(\ell+1)(2\ell+1)}$. In the third step the coordinates $\{x^3,x^4\}$ adapted to the orientation of $S_t$ was chosen such that the triplet with the right-handed orientation consists of the coordinate basis $\partial/\partial x^3$, $\partial/\partial x^4$, and the normal vector $n^a_{S_t}$ of $S_t$ in 3-dimensional $\Sigma$. In the fourth step the identity $ e^i_I=\frac{2}{\kappa\beta}\{A^i_I,V\}$ was used. In the fifth step a partition ${\cal P}_{\epsilon'} (S_t)$ of $S$ adapted to the coordinates $\{x^3,x^4\}$ into boxes $\Box$ with parameter area $\epsilon'^2$ was implemented, and $e^t_I(\Box)$ are the edges starting from $v^t(\Box)$ along $I$-th coordinate lines with positive orientation and parameter length $\epsilon'$ (see Fig. \ref{partition}). In the sixth step the identity $\epsilon_{ijk}=-6\,\left[\chi(\ell)\right]^{-2}{\rm Tr}_{\ell}(\tau_i\tau_j\tau_k)$ was used, here $\tau_i=-\frac{{\rm i}}{2}\sigma_i$ (with $\sigma_i$ being the Pauli matrices) and ${\rm Tr}_\ell$ denotes the trace in the representation $\pi_\ell$ with spin $\ell$. In the last step we have used the identity
\begin{align}
\pi_\ell(h_{e^t_I(\Box)})\left\{\pi_\ell(h_{e^t_I(\Box)}^{-1}),V_{v^t(\Box)}\right\}&=-\epsilon'\left\{A^j_I\,\pi_\ell(\tau_j),V_{v^t(\Box)}\right\}+O({\epsilon'}^{2})\,,
\end{align} 
where $h_{e^t_I(\Box)}\equiv h_{e^t_I(\Box)}(A)$ indicates the holonomies of connection along edges $e^t_I(\Box)$, and $\hat{V}_{v^t(\Box)}$ denotes the volume operator corresponding to the classical function $V_{v^t(\Box)}$ of a 3-dimensional region $R_{v^t(\Box)}$ containing $v^t(\Box)$, and $R_{v^t(\Box)}\rightarrow v^t(\Box)$ as the limit $\epsilon'\rightarrow 0$.

To quantize $\tilde{E}^{\rm Alt}_i(S_t)$ expressed in Eq. \eqref{class-alt-reg-i} in a certain manner such that its quantum version is consistent with the fundamental flux operator $\hat{\tilde{E}}^{\rm Fun}_i(S_t)$ in Eq. \eqref{fun-action-i}, we replace $V$ by its operator version $\hat{V}$, holonomies by holonomy operators (since the holonomy operator acts as a multiplication operator, we also omit the hat for simplification of notation), and the Poisson bracket by $1/({\rm i}\hbar)$ times the commutator. Then we obtain the alternative flux operator $\hat{\tilde{E}}^{\rm Alt}_i(S_t)$ after removing the regulator $\epsilon'$ by taking the limit $\epsilon'\rightarrow0$. In this paper we only consider the volume operator defined in \cite{Ashtekar:1997fb,Thiemann:1996au}, which was used to define a Hamiltonian constraint operator in LQG \cite{Thiemann:1996aw}, and it is given by
\begin{align}\label{volume-operator}
\hat{V}_\gamma=\sum_{v\in V(\gamma)}\sqrt{\left|\hat{Q}_v\right|}=\sum_{v\in V(\gamma)}\sqrt{\left|\kappa_{\rm reg}\frac{{\rm i}\ell_{\rm p}^6\,\beta^3}{32}\sum_{I<J<K}\varsigma(e_I,e_J,e_K)\,\hat{q}_{IJK}\right|}\,,
\end{align}
where the sum $\sum_{I<J<K}$ is over all triples $(e_I,e_J,e_K)$ of edges at the vertex $v\in V(\gamma)$ for a given graph $\gamma$, $\kappa_{\rm reg}$ denotes the regularization constant coming from averaging over the relevant background structures in the regularization procedure of the volume operator \cite{Ashtekar:1997fb}, $\varsigma(e_I,e_J,e_K)\equiv{\rm sgn}\left[\det\left(\dot{e}_I(0),\dot{e}_J(0),\dot{e}_K(0)\right)\right]$ takes the values of $0$, $+1$ and $-1$, corresponding to whether the determinant of the matrix formed by the tangents of the three edges at $v$ in that sequence is zero, positive, or negative, and
\begin{align}
\hat{q}_{IJK}\equiv -4{\rm i}\epsilon_{ijk}J^i_{e_I}J^j_{e_J}J^k_{e_K}\,.
\end{align}
Due to the factor $\varsigma(e_I,e_J,e_K)$, the volume operator $\hat{V}$ vanishes on the linearly dependent triplets. Since Eq. \eqref{class-alt-reg-i} contains the volume $V_{v^t(\Box)}$, in order to get an alternative flux operator whose action on the edges of type up or down with respect to the surface $S_t$ takes the similar result as that of the fundamental flux operator, the holonomies involving $e^t_3(\Box)$ and $e^t_4(\Box)$ should be arranged to the right-hand side of $V_{v^t(\Box)}$.  Under the above considerations and noticing that holonomies commute with each other classically, the alternative flux is arranged as the following ordering
\begin{widetext}
\begin{align}\label{class-flux}
\tilde{E}^{\rm Alt}_i(S_t)&=-24\left[\kappa\,\beta\,\chi(\ell)\right]^{-2}\lim_{\epsilon'\rightarrow0}\sum_{\Box\in\,{\cal P}_{\epsilon'}(S_t)}{[\pi_\ell(\tau_i)]^B}_C{[\pi_\ell(h_{e^t_4(\Box)})]^C}_D\left\{{[\pi_\ell(h_{e^t_3(\Box)}^{-1})]^A}_B,V_{v^t(\Box)}\right\}{\cal S}\, \left\{V_{v^t(\Box)},{[\pi_\ell(h_{e^t_4(\Box)}^{-1})]^D}_E\right\}{[\pi_\ell(h_{e^t_3(\Box)})]^E}_A\,,
\end{align}
\end{widetext}
where the indices $A,B,\cdots=-\ell,-\ell+1,\cdots,\ell$, the upper (or former) indices are the row indices while the lower (or later) are the column indices in the matrix elements. It is shown that ${\cal S}$ can be identified with the sign that appears inside the absolute value under the square roots in the definition of the volume \cite{Giesel:2005bk,Giesel:2005bm}. Finally, the quantum version of the alternative flux function $\tilde{E}^{\rm Alt}_i(S)$ in Eq. \eqref{class-flux} can be written as
\begin{widetext}
\begin{align}\label{Alt-flux-operator-def}
 \hat{\tilde{E}}^{\rm Alt}_i(S_t)&=24\left[\ell_{\rm p}^2\,\beta\,\chi(\ell)\right]^{-2}\lim_{\epsilon'\rightarrow0}\sum_{\Box\in\,{\cal P}_{\epsilon'}(S_t)}{[\pi_\ell(\tau_i)]^B}_C{[\pi_\ell(h_{e^t_4(\Box)})]^C}_D\left[{[\pi_\ell(h_{e^t_3(\Box)}^{-1})]^A}_B,\hat{V}_{v^t(\Box)}\right]\hat{\cal S}\, \left[\hat{V}_{v^t(\Box)},{[\pi_\ell(h_{e^t_4(\Box)}^{-1})]^D}_E\right]{[\pi_\ell(h_{e^t_3(\Box)})]^E}_A\,,
\end{align}
\end{widetext}
where $\hat{\cal S}$ is the sign operator defined by $\hat{Q}_{v^t(\Box)}=:\hat{V}_{v^t(\Box)}\hat{\cal S}\,\hat{V}_{v^t(\Box)}$ \cite{Giesel:2005bk,Giesel:2005bm}. To compute the action of the alternative flux operator on a spin network state, the graphical calculus based on the original Brink graphical method is adopted (see e.g. \cite{Yang:2015wka} for reference). In practical calculation, it is convenient to introduce the spherical tensors $\tau_\mu$ $(\mu=0,\pm1)$, corresponding to $\tau_i$ $(i=1,2,3)$, defined by
\begin{align}
\tau_0:=\tau_3,\qquad \tau_{\pm1}:=\mp\frac{1}{\sqrt{2}}\left(\tau_1\pm\tau_2\right)\,.
\end{align}
Then the alternative flux operator defined by $\tau_\mu$ is given by
\begin{widetext}
\begin{align}\label{Alt-flux-tau-def}
\hat{\tilde{E}}^{\rm Alt}_\mu(S_t)&=24\left[\ell_{\rm p}^2\,\beta\,\chi(\ell)\right]^{-2}\lim_{\epsilon'\rightarrow0}\sum_{\Box\in\,{\cal P}_{\epsilon'}(S_t)}{[\pi_\ell(\tau_\mu)]^B}_C{[\pi_\ell(h_{e^t_4(\Box)})]^C}_D\left[{[\pi_\ell(h_{e^t_3(\Box)}^{-1})]^A}_B,\hat{V}_{v^t(\Box)}\right]\hat{\cal S}\, \left[\hat{V}_{v^t(\Box)},{[\pi_\ell(h_{e^t_4(\Box)}^{-1})]^D}_E\right]{[\pi_\ell(h_{e^t_3(\Box)})]^E}_A\,.
\end{align}
\end{widetext}

Let us now consider the spin network state corresponding to the original edge $e$ in Fig. \ref{fig-up} (a). Assigning a spin $j$ to the original edge ${}^{\sst (u)}\!e$ in Fig. \ref{fig-up}, the spin network state corresponding to this edge is ${[\pi_j(h_{{}^{\sst (u)}\!e})]^m}_n$. Partition of ${}^{\sst (u)}\!e$ at $t$ into two edges ${}^{\sst (u)}\!e^t_1$ and ${}^{\sst (u)}\!e^t_2$ induces a spin network state associated to two edges ${}^{\sst (u)}\!e^t_1$ and ${}^{\sst (u)}\!e^t_2$ in Fig. \ref{fig-up} (b) as (the derivation in graphical calculus will be given below)
\begin{align}\label{state-alg}
 \left|{\left({}^{\sst (u)}\!\beta_t\right)^{J=0}}_{M=0}\right\rangle&\equiv{[\pi_j(h_e)]^m}_n=\sqrt{2j+1}\,{\left(i^{J=0}_{\tilde{v}_t}\right)_{m_1m_2}}^{M=0}\,{[\pi_j(h_{{}^{\sst (u)}\!e^t_1})]^{m_1}}_{n_1}{[\pi_j(h_{{}^{\sst (u)}\!e^t_2})]^{m_2}}_{n_2}\,,
\end{align}
where ${\left(i^{J=0}_{\tilde{v}_t}\right)_{m_1m_2}}^{M=0}\equiv\langle J=0,M=0|jm_1;jm_2\rangle$ denotes the normalized gauge-invariant intertwiner at $\tilde{v}_t$, which describes the coupling of two angular momenta $j, j$ with the magnetic quantum numbers $m_1, m_2$ to a total angular momentum $J=0$ with the magnetic quantum number $M=0$. Notice that $\hat{\tilde{E}}^{\rm Alt}_\mu(S_t)$ involves the volume operator $\hat{V}$ which has non-trivial action only on the states containing at least one non-coplanar trivalent or multivalent vertex. Therefore, in order to obtain a non-trial action of the alternative flux operator on ${\left(i^{J=0}_{\tilde{v}_t}\right)_{m_1m_2}}^{M=0}$, the partition ${\cal P}_{\epsilon'}(S_t)$ of $S_t$ should be graph-dependently chosen in such a way that the graph $\gamma$ intersects at least one $\Box\in\,{\cal P}_{\epsilon'}(S_t)$ at $\tilde{v}_t=v^t(\Box)$. For the partition adapted to $\gamma$, the sum over $\Box\in\,{\cal P}_{\epsilon'}(S_t)$ in $\hat{\tilde{E}}^{\rm Alt}_\mu(S_t)$ reduces to the only one box $\Box$ that intersects $\gamma$ at $\tilde{v}_t$, and we will omit the only box $\Box$ for simplifications. Since the volume operator vanishes the co-planar vertices, only one term in the commutator in the alternative flux operator $\hat{\tilde{E}}^{\rm Alt}_\mu(S_t)$ has nontrivial contribution. Hence $\hat{\tilde{E}}^{\rm Alt}_\mu(S_t)$ defined in Eq. \eqref{Alt-flux-tau-def} acts on the state $\left|{\left({}^{\sst (u)}\!\beta_t\right)^{J=0}}_{M=0}\right\rangle$ yields
\begin{widetext}
\begin{align}\label{Alt-tau-sim-action}
 \hat{\tilde{E}}^{\rm Alt}_\mu(S_t)\, \left|{\left({}^{\sst (u)}\!\beta_t\right)^{J=0}}_{M=0}\right\rangle&=24\left[\ell_{\rm p}^2\,\beta\,\chi(\ell)\right]^{-2}\lim_{\epsilon'\rightarrow0}{[\pi_\ell(\tau_\mu)]^B}_C{[\pi_\ell(h_{e^t_4})]^C}_D{[\pi_\ell(h_{e^t_3}^{-1})]^A}_B\hat{V}_{\tilde{v}_t}\hat{\cal S}\, \hat{V}_{\tilde{v}_t}{[\pi_\ell(h_{e^t_4}^{-1})]^D}_E{[\pi_\ell(h_{e^t_3})]^E}_A\, \left|{\left({}^{\sst (u)}\!\beta_t\right)^{J=0}}_{M=0}\right\rangle\notag\\
  &=24\left[\ell_{\rm p}^2\,\beta\,\chi(\ell)\right]^{-2}\lim_{\epsilon'\rightarrow0}{[\pi_\ell(\tau_\mu)]^B}_C{[\pi_\ell(h_{e^t_4})]^C}_D{[\pi_\ell(h_{e^t_3}^{-1})]^A}_B\hat{Q}_{\tilde{v}_t}{[\pi_\ell(h_{e^t_4}^{-1})]^D}_E{[\pi_\ell(h_{e^t_3})]^E}_A \, \left|{\left({}^{\sst (u)}\!\beta_t\right)^{J=0}}_{M=0}\right\rangle\,.
\end{align}
\end{widetext}

We now compute Eq. \eqref{Alt-tau-sim-action} by the graphical method. Note that the initial spin network state corresponding to Fig. \ref{fig-up} (a)  and its induced spin network state corresponding to \ref{fig-up} (b) in the algebraic Eq. \eqref{state-alg} are related by (see ref. \cite{Yang:2015wka} for details)
\begin{widetext}
\begin{align}\label{init-state-up}
 \left|{\left({}^{\sst (u)}\!\beta_t\right)^{J=0}}_{M=0}\right\rangle&=\makeSymbol{
\includegraphics[height=3.8cm]{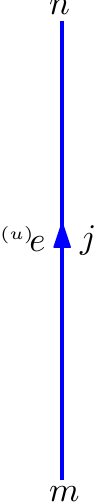}}=\makeSymbol{
\includegraphics[height=3.8cm]{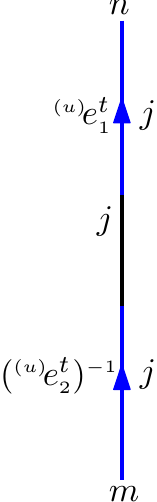}}=\makeSymbol{
\includegraphics[height=3.8cm]{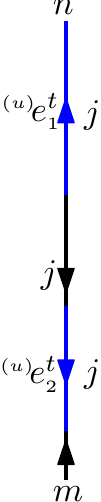}}=\sqrt{2j+1}\makeSymbol{
\includegraphics[height=3.8cm]{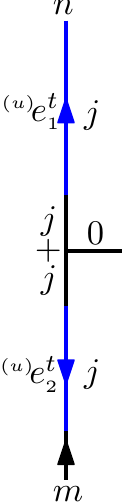}}=\sqrt{2j+1}\left[\makeSymbol{
\includegraphics[height=2.5cm]{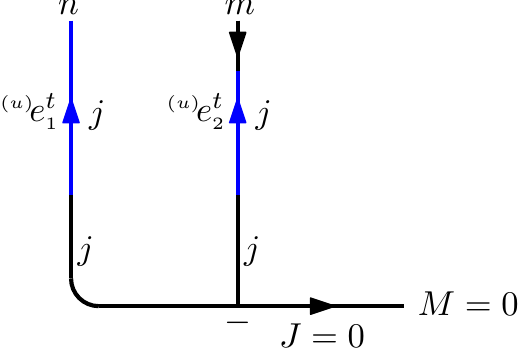}}\right]\,,
\end{align}
\end{widetext}
where in the second and the third steps we have used Eqs. (A.36) and (3.17) in \cite{Yang:2015wka}, respectively, in the four step we have used the graphical rule (see e.g. Eq. (A.47) in \cite{Yang:2015wka})
\begin{align}\label{graph-rule-triagle}
 \makeSymbol{
\includegraphics[width=2cm]{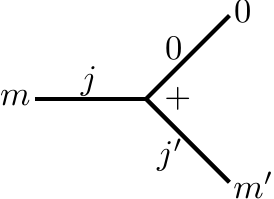}}=\frac{\delta_{j,j'}}{\sqrt{2j+1}}\;\makeSymbol{
\includegraphics[width=2cm]{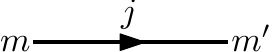}}\,.
\end{align}
The action of $\hat{\tilde{E}}^{\rm Alt}_\mu(S_t)$ on the state $\left|{\left({}^{\sst (u)}\!\beta_t\right)^{J=0}}_{M=0}\right\rangle$ consists of the following four steps.

In the first step, we consider the action of the two holonomies on the most right-hand side of $\hat{\tilde{E}}^{\rm Alt}_\mu(S_t)$ in Eq. \eqref{Alt-tau-sim-action}. The alternative flux operator acts on $\left|{\left({}^{\sst (u)}\!\beta_t\right)^{J=0}}_{M=0}\right\rangle$ by attaching two additional edges $e^t_3$ and $e^t_4$ to the edge $e$ (and to ${}^{\sst (u)}\!e^t_1$ and ${}^{\sst (u)}\!e^t_2$). Notice that the holonomy operator acts as a multiplication operator. Thus the two matrix elements of holonomies can be represented as
\begin{widetext}
\begin{align}
 {[\pi_\ell(h_{e^t_4}^{-1})]^D}_E{[\pi_\ell(h_{e^t_3})]^E}_A&=\makeSymbol{
\includegraphics[height=2.4cm]{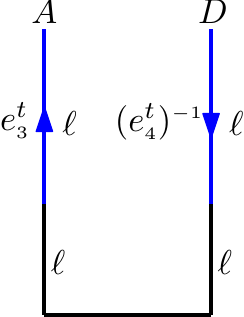}}=(2\ell+1)\makeSymbol{
\includegraphics[height=2.5cm]{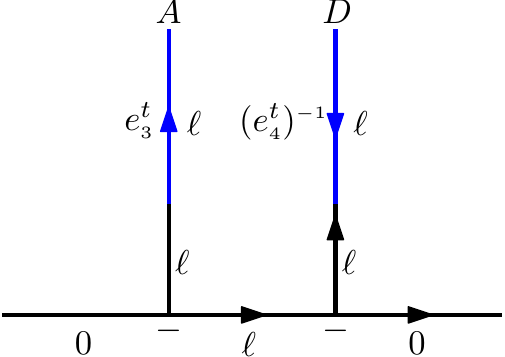}}=(2\ell+1)\makeSymbol{
\includegraphics[height=2.5cm]{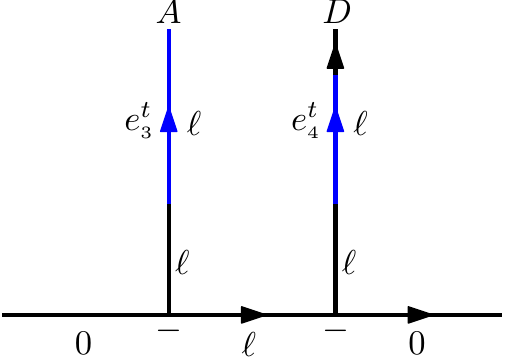}}\,,
\end{align}
\end{widetext}
where we have used Eq. \eqref{graph-rule-triagle} in the second step. Then the two holonomies in $\hat{\tilde{E}}^{\rm Alt}_\mu(S_t)$ on the most right-hand side act on $\left|{\left({}^{\sst (u)}\!\beta_t\right)^{J=0}}_{M=0}\right\rangle$ as 
\begin{widetext}
\begin{align}\label{action:first step}
\hat{\tilde{E}}^{\rm Alt}_\mu(S_t)\,\left|{\left({}^{\sst (u)}\!\beta_t\right)^{J=0}}_{M=0}\right\rangle
&=24\left[\ell_{\rm p}^2\,\beta\,\chi(\ell)\right]^{-2}\lim_{\epsilon'\rightarrow0}\sqrt{2j+1}\sqrt{2\ell+1}\,{[\pi_\ell(\tau_\mu)]^B}_C{[\pi_\ell(h_{e^t_4})]^C}_D{[\pi_\ell(h_{e^t_3}^{-1})]^A}_B\hat{Q}_{\tilde{v}_t}\notag\\
&\hspace{3cm}\times\left[\sqrt{2\ell+1}\makeSymbol{
\includegraphics[height=2.5cm]{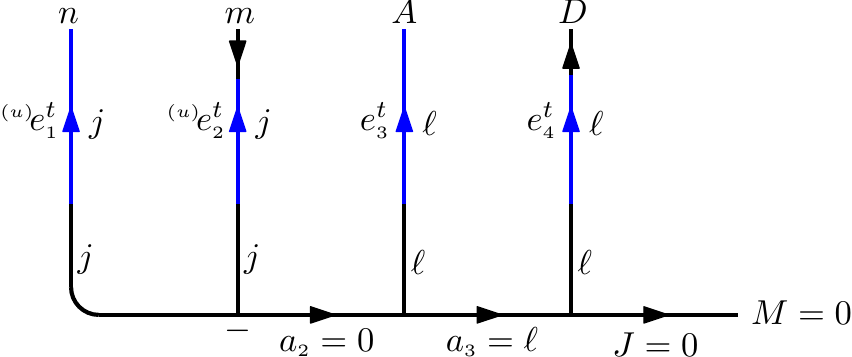}}\right]\,.
\end{align}
\end{widetext}

In the second step, we consider the action of $\hat{Q}_{\tilde{v}_t}$. Notice that
\begin{align}\label{Q-sum}
 \hat{Q}_{\tilde{v}_t}&=\kappa_{\rm reg}\frac{{\rm i}\ell_{\rm p}^6\,\beta^3}{32}\sum_{I<J<K}\epsilon(e_I,e_J,e_K)\hat{q}_{IJK}\notag\\
 &=\kappa_{\rm reg}\frac{{\rm i}\ell_{\rm p}^6\,\beta^3}{32}\left[\epsilon(1,3,4)\hat{q}_{134}+\epsilon(2,3,4)\hat{q}_{234}\right]\notag\\
 &=\kappa_{\rm reg}\frac{{\rm i}\ell_{\rm p}^6\,\beta^3}{32}\left(\hat{q}_{134}-\hat{q}_{234}\right)
\end{align}
is gauge invariant, and thus it only changes the intermediate couplings ${\vec{a}}$ in the intertwiner associated to the spin network state $\,\left|{\left({}^{\sst (u)}\!\beta_t\right)^{J=0}}_{M=0}\right\rangle$. Denote the gauge-invariant intertwiner of $\,\left|{\left({}^{\sst (u)}\!\beta_t\right)^{J=0}}_{M=0}\right\rangle$ at $\tilde{v}_t$ by $|a_2=0,a_3=\ell,J=0\rangle$. Then the action of $\hat{Q}_{\tilde{v}_t}$ on $|a_2=0,a_3=\ell,J=0\rangle$ can be linearly expanded by $|a'_2,a'_3=\ell,J=0\rangle$ as
\begin{widetext}
\begin{align}
\hat{Q}_{\tilde{v}_t}|a_2=0,a_3=\ell,J=0\rangle&=\sum_{a'_2}\langle a'_2,a'_3=\ell,J=0|\hat{Q}_{\tilde{v}_t}|a_2=0,a_3=\ell,J=0\rangle\;|a'_2,a'_3=\ell,J=0\rangle\notag\\
&=\sum_{a'_2}\kappa_{\rm reg}\frac{{\rm i}\ell_{\rm p}^6\,\beta^3}{32}\left[1-(-1)^{a'_2}\right]\langle a'_2,a'_3=\ell,J=0|\hat{q}_{134}|a_2=0,a_3=\ell,J=0\rangle\;|a'_2,a'_3=\ell,J=0\rangle\notag\\
&=\sum_{a'_2}\kappa_{\rm reg}\frac{{\rm i}\ell_{\rm p}^6\,\beta^3}{4\sqrt{3}}\sqrt{j(j+1)}\sqrt{\ell(\ell+1)}\,\delta_{a'_2,1}\;|a'_2,a'_3=\ell,J=0\rangle\notag\\
&=\kappa_{\rm reg}\frac{{\rm i}\ell_{\rm p}^6\,\beta^3}{4\sqrt{3}}\sqrt{j(j+1)}\sqrt{\ell(\ell+1)}\;|a'_2=1,a'_3=\ell,J=0\rangle\,,
\end{align}
\end{widetext}
where we have used the formula of the matrix elements of $\hat{q}_{IJK}$ (see, e.g. Eqs. (4.35) and (4.36) in \cite{Brunnemann:2004xi,Yang:2015wka} for details). Then Eq. \eqref{action:first step} yields
\begin{widetext}
\begin{align}\label{action:second step}
 \hat{\tilde{E}}^{\rm Alt}_\mu(S_t)\,\left|{\left({}^{\sst (u)}\!\beta_t\right)^{J=0}}_{M=0}\right\rangle
&=24\left[\ell_{\rm p}^2\,\beta\,\chi(\ell)\right]^{-2}\lim_{\epsilon'\rightarrow0}\kappa_{\rm reg}\frac{{\rm i}\ell_{\rm p}^6\,\beta^3}{4\sqrt{3}}\,\chi(j)\,\chi(\ell)\,{[\pi_\ell(\tau_\mu)]^B}_C{[\pi_\ell(h_{e^t_4})]^C}_D{[\pi_\ell(h_{e^t_3}^{-1})]^A}_B\notag\\
&\hspace{3cm}\times\left[\sqrt{3(2\ell+1)}\makeSymbol{
\includegraphics[height=2.5cm]{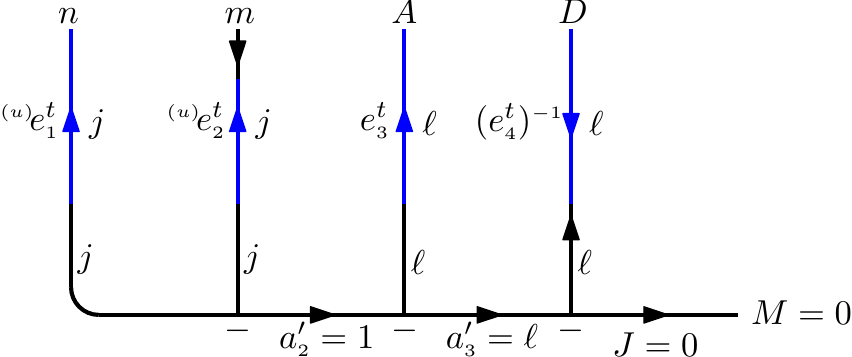}}\right]\,.
\end{align}
\end{widetext}

In the third step, we implement the action of the remaining holonomies. The action only involves the contractions of holonomy and its inverse, which is given by
\begin{align}
 {[\pi_\ell(h)]^A}_B{[\pi_\ell(h^{-1})]^B}_C=\delta^A_C\,.
\end{align}
Hence the result of this action yields
\begin{widetext}
\begin{align}\label{action:third step}
 \hat{\tilde{E}}^{\rm Alt}_\mu(S_t)\,\left|{\left({}^{\sst (u)}\!\beta_t\right)^{J=0}}_{M=0}\right\rangle&=24\left[\ell_{\rm p}^2\,\beta\,\chi(\ell)\right]^{-2}\lim_{\epsilon'\rightarrow0}\kappa_{\rm reg}\frac{{\rm i}\ell_{\rm p}^6\,\beta^3}{4\sqrt{3}}\,\chi(j)\,\chi(\ell)\,{[\pi_\ell(\tau_\mu)]^B}_C\left[\sqrt{3(2\ell+1)}\makeSymbol{
\includegraphics[height=2.5cm]{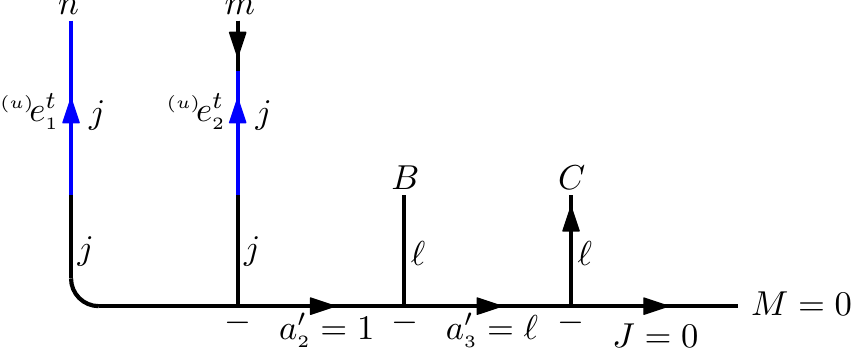}}\right]\,.
 \end{align}
\end{widetext}
 
In the last step,  we deal with the action of ${[\pi_\ell(\tau_\mu)]^B}_C$, which involves the contraction of ${[\pi_\ell(\tau_\mu)]^B}_C$ with the state it acts. Notice that the matrix element of ${[\pi_\ell(\tau_\mu)]^B}_C$ can be expressed by \cite{Yang:2015wka}
\begin{align}
{[\pi_\ell(\tau_\mu)]^B}_C&={\rm i}\,\chi(\ell)\,\makeSymbol{
\includegraphics[width=1.5cm]{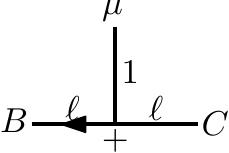}}\,.
\end{align}
Then we can write down the action as
\begin{widetext}
 \begin{align}\label{Alt-flux-action}
 \hat{\tilde{E}}^{\rm Alt}_\mu(S_t)\,\left|{\left({}^{\sst (u)}\!\beta_t\right)^{J=0}}_{M=0}\right\rangle&=-24\left[\ell_{\rm p}^2\,\beta\,\chi(\ell)\right]^{-2}\lim_{\epsilon'\rightarrow0}\kappa_{\rm reg}\frac{\ell_{\rm p}^6\,\beta^3}{4\sqrt{3}}\,\chi(j)\,\left[\chi(\ell)\right]^2\,\left[\sqrt{3(2\ell+1)}\makeSymbol{
\includegraphics[height=2.5cm]{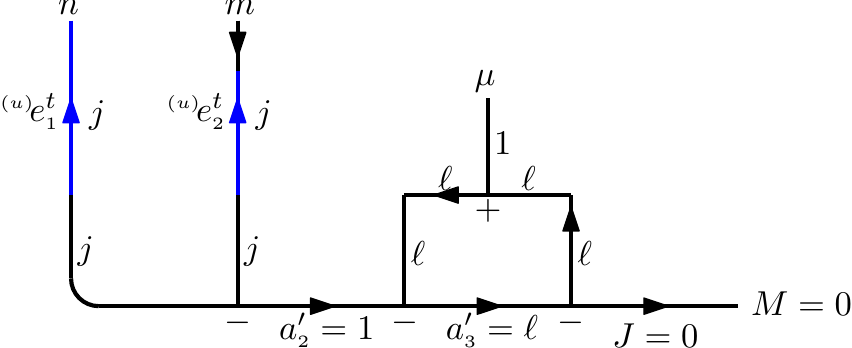}}\right]\notag\\
&=-2\kappa_{\rm reg}\ell_{\rm p}^2\,\beta\,\chi(j)\lim_{\epsilon'\rightarrow0}\left[\sqrt{3}\makeSymbol{
\includegraphics[height=2.5cm]{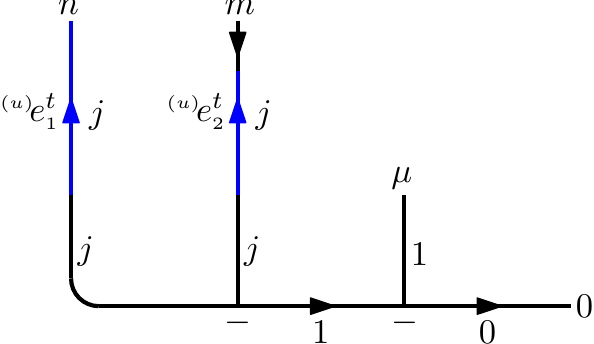}}\right]=-2\kappa_{\rm reg}\ell_{\rm p}^2\,\beta\,\chi(j)\makeSymbol{
\includegraphics[height=3.8cm]{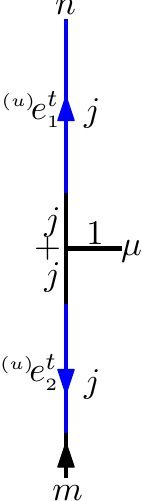}}\,,
\end{align}
\end{widetext}\noindent
where in the second step we have used
\begin{align}
 \makeSymbol{
\includegraphics[width=3cm]{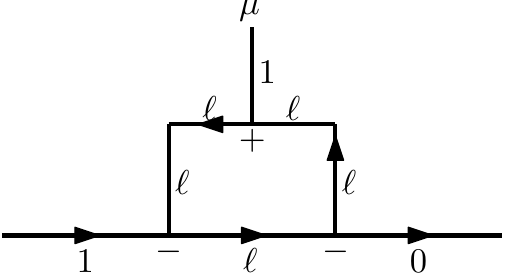}}&=\frac{1}{\sqrt{2\ell+1}}\makeSymbol{
\includegraphics[width=2.1cm]{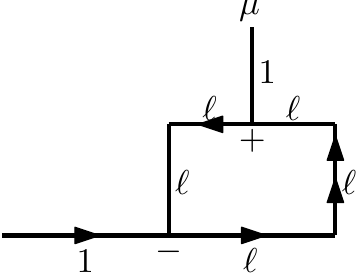}}=\frac{1}{\sqrt{2\ell+1}}\makeSymbol{
\includegraphics[width=2.1cm]{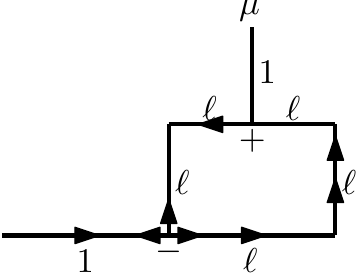}}=\frac{1}{\sqrt{2\ell+1}}\makeSymbol{
\includegraphics[width=3cm]{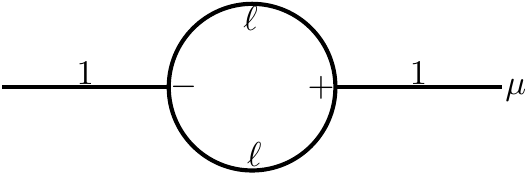}}\notag\\
&=\frac{1}{3\sqrt{2\ell+1}}\makeSymbol{
\includegraphics[width=1cm]{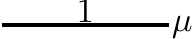}}=\frac{1}{\sqrt{3(2\ell+1})}\makeSymbol{
\includegraphics[width=2cm]{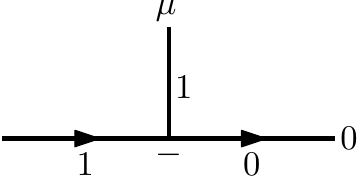}}\,,
\end{align}
and the trivial limit was taken in the third step.

On the other hand, the action of the fundamental flux operator $\hat{\tilde{E}}^{\rm Fun}_\mu(S_t)$ corresponding to $\hat{\tilde{E}}^{\rm Fun}_i(S_t)$ in Eq. \eqref{fun-action-i} on a cylindrical function is given by
\begin{align}\label{Fun-flux-tau-def}
 \hat{\tilde{E}}^{\rm Fun}_\mu(S_t)\,\cdot f_\gamma&=\frac{\ell_{\rm p}^2\,\beta}{2}\sum_{v\in \gamma\cap S_t}\sum_{b(e)=v}\varrho(e,S_t)J^\mu_e\,\cdot f_\gamma\,.
\end{align}
On the same state $\left|{\left({}^{\sst (u)}\!\beta_t\right)^{J=0}}_{M=0}\right\rangle$, its action can be simplified algebraically as
\begin{align}\label{Fun-action-simp}
 \hat{\tilde{E}}^{\rm Fun}_\mu(S_t)\,\left|{\left({}^{\sst (u)}\!\beta_t\right)^{J=0}}_{M=0}\right\rangle&=\frac{\ell_{\rm p}^2\,\beta}{2}\sum_{v\in \gamma\cap S_t}\sum_{b(e)=v}\varrho(e,S_t)J^\mu_e\,\left|{\left({}^{\sst (u)}\!\beta_t\right)^{J=0}}_{M=0}\right\rangle\notag\\
 &=\frac{\ell_{\rm p}^2\,\beta}{2}\left[\varrho({}^{\sst (u)}\!e^t_1,S_t)J^\mu_{{}^{\sst (u)}\!e^t_1}+\varrho({}^{\sst (u)}\!e^t_2,S_t)J^\mu_{{}^{\sst (u)}\!e^t_2}\right]\,\left|{\left({}^{\sst (u)}\!\beta_t\right)^{J=0}}_{M=0}\right\rangle\notag\\
 &=\ell_{\rm p}^2\,\beta\;\varrho({}^{\sst (u)}\!e^t_2,S_t)J^\mu_{{}^{\sst (u)}\!e^t_2}\,\left|{\left({}^{\sst (u)}\!\beta_t\right)^{J=0}}_{M=0}\right\rangle\notag\\
 &=-\ell_{\rm p}^2\,\beta\;J^\mu_{{}^{\sst (u)}\!e^t_2}\,\left|{\left({}^{\sst (u)}\!\beta_t\right)^{J=0}}_{M=0}\right\rangle\,,
\end{align}
where in the third step we used the fact that, for the state $\left|{\left({}^{\sst (u)}\!\beta_t\right)^{J=0}}_{M=0}\right\rangle$, gauge invariance at $\tilde{v}_t$ implies $\left(J_{{}^{\sst (u)}\!e^t_1}^\mu+J_{{}^{\sst (u)}\!e^t_2}^\mu\right)\,\left|{\left({}^{\sst (u)}\!\beta_t\right)^{J=0}}_{M=0}\right\rangle=0$. In graphical calculus the action in Eq. \eqref{Fun-action-simp} can be written as (see \cite{Yang:2015wka})
\begin{align}\label{Fun-flux-action}
  \hat{\tilde{E}}^{\rm Fun}_\mu(S_t)\,\left|{\left({}^{\sst (u)}\!\beta_t\right)^{J=0}}_{M=0}\right\rangle&=-\ell_{\rm p}^2\,\beta\,\chi(j)\sqrt{2j+1}\makeSymbol{
\includegraphics[height=2.5cm]{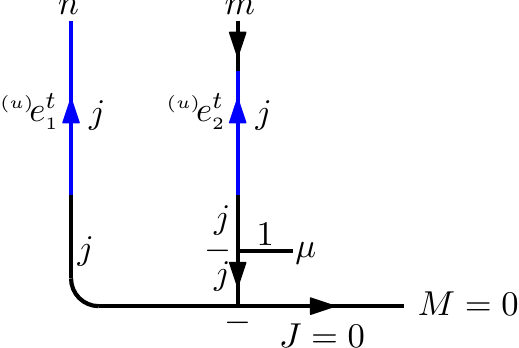}}=-\ell_{\rm p}^2\,\beta\,\chi(j)\left[\sqrt{3}\makeSymbol{
\includegraphics[height=2.5cm]{figures/SNF-vertex-5}}\right]\notag\\
&=-\ell_{\rm p}^2\,\beta\,\chi(j)\makeSymbol{
\includegraphics[height=3.8cm]{figures/SNF-vertex-acted}}\,,
\end{align}
where in the second step we have used the identity Eq. (4.14) in \cite{Yang:2015wka}.

To summarize, the actions of the two flux operators on $\left|{\left({}^{\sst (u)}\!\beta_t\right)^{J=0}}_{M=0}\right\rangle$ in Eqs. \eqref{Alt-flux-action} and \eqref{Fun-flux-action} can be written as
\begin{align}\label{Alt-action-graph-up}
 \hat{\tilde{E}}^{\rm Fun/Alt}_\mu(S_t)\,\left|{\left({}^{\sst (u)}\!\beta_t\right)^{J=0}}_{M=0}\right\rangle=\hat{\tilde{E}}^{\rm Fun/Alt}_\mu(S_t)\,\makeSymbol{
\includegraphics[height=3.8cm]{figures/SNF-origin-edge}}&=\hat{\tilde{E}}^{\rm Fun/Alt}_\mu(S_t)\,\sqrt{2j+1}\makeSymbol{
\includegraphics[height=3.8cm]{figures/SNF-origin-edge-3}}=-\alpha^{\rm Fun/Alt}\ell_{\rm p}^2\,\beta\,\chi(j)\,\makeSymbol{
\includegraphics[height=3.8cm]{figures/SNF-vertex-acted}} \,,
\end{align}
where the factor $\alpha^{\rm Fun/Alt}$ takes $1$/$(2\kappa_{\rm reg})$ for the fundamental/alternative flux operator.

\begin{figure}[t]
\centering
\begin{minipage}[b]{0.2\textwidth}
\begin{tabular}{c}
\includegraphics[]{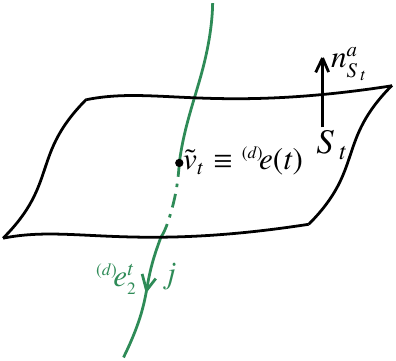}\\
(a)
\end{tabular}
\end{minipage}
\hspace{0.3cm}
\begin{minipage}[b]{0.2\textwidth}
\begin{tabular}{c}
\includegraphics[]{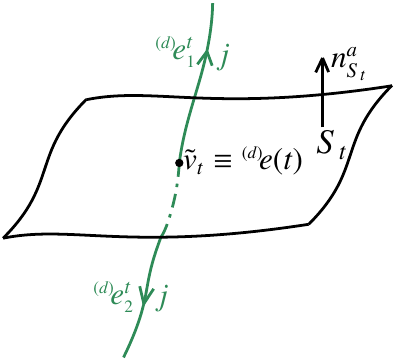}\\
(b)
\end{tabular}
\end{minipage}
\caption{(a) A edge ${}^{\sst (d)}\!e$ intersects a surface $S_t$ intersects at $\tilde{v}_t\equiv {}^{\sst (d)}\!e(t)$. (b) The original edge ${}^{\sst (d)}\!e$ is partitioned by $S_t$ into two edges ${}^{\sst (d)}\!e^t_1$ and ${}^{\sst (d)}\!e^t_2$ starting from $\tilde{v}_t$.}
\label{fig-down}
\end{figure}

Now we consider another case different from Fig. \ref{fig-up}, where an edge in a graph $\gamma$, ${}^{\sst (d)}\!e:[0,1]\rightarrow \Sigma$, isolated intersects a surface denoted by $S_t$ at an additional vertex $\tilde{v}_t\equiv {}^{\sst (d)}\!e(t)$ with ${}^{\sst (d)}\!\dot{e}^a(t)n_a^{S_t}<0$, and it is subdivided into two edges ${}^{\sst (d)}\!e^t_1$ and ${}^{\sst (d)}\!e^t_2$ starting from $\tilde{v}_t$ (see Fig. \ref{fig-down}). Similarly, the initial spin network state $\left|{\left({}^{\sst (d)}\!\beta_t\right)^{J=0}}_{M=0}\right\rangle$ corresponding to Fig. \ref{fig-down} (a)  and its induced spin network state corresponding to Fig. \ref{fig-down} (b) are related by
\begin{align}\label{init-state-down}
 \left|{\left({}^{\sst (d)}\!\beta_t\right)^{J=0}}_{M=0}\right\rangle&=\makeSymbol{
\includegraphics[height=3.8cm]{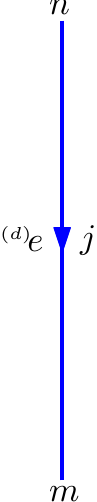}}=\makeSymbol{
\includegraphics[height=3.8cm]{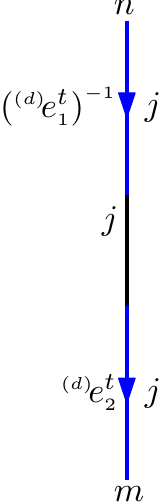}}=\makeSymbol{
\includegraphics[height=3.8cm]{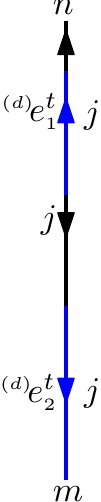}}=\sqrt{2j+1}\makeSymbol{
\includegraphics[height=3.8cm]{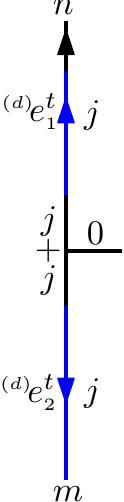}}\,.
\end{align}

Similar to the above calculations, we obtain the actions of the two flux operators $\hat{\tilde{E}}^{\rm Fun/Alt}_\mu(S_t)$ on $\left|{\left({}^{\sst (d)}\!\beta_t\right)^{J=0}}_{M=0}\right\rangle$ as
\begin{align}\label{Alt-action-graph-down}
 \hat{\tilde{E}}^{\rm Fun/Alt}_\mu(S_t)\,\left|{\left({}^{\sst (d)}\!\beta_t\right)^{J=0}}_{M=0}\right\rangle=\hat{\tilde{E}}^{\rm Fun/Alt}_\mu(S_t)\,\makeSymbol{
\includegraphics[height=3.8cm]{figures/SNF-origin-edge-down}}=\hat{\tilde{E}}^{\rm Fun/Alt}_\mu(S_t)\,\sqrt{2j+1}\makeSymbol{
\includegraphics[height=3.8cm]{figures/SNF-origin-edge-down-3}}=-\alpha^{\rm Fun/Alt}\ell_{\rm p}^2\,\beta\,\chi(j)\makeSymbol{
\includegraphics[height=3.8cm]{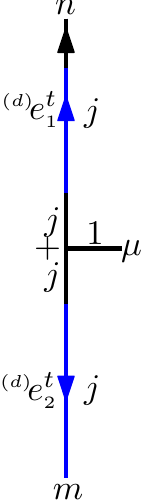}}\,.
\end{align}
Thus, the coefficients in front of the resulting spin network states in the two cases are the same.

\subsection{The alternative flux operator as a limitation and the consistency check}
let us consider whether the alternative flux operator $\hat{\tilde{E}}^{\rm Alt}_\mu(S_t)$ defined in Eq. \eqref{Alt-flux-tau-def} can be consistent with the fundamental flux operator $\hat{\tilde{E}}^{\rm Fun}_\mu(S_t)$ in Eq. \eqref{Fun-flux-tau-def} for all the cases corresponding to the relation between a surface $S_t$ and a graph $\gamma$. The first case is that the intersection points locate at interior points of the edges of $\gamma$, which is the case discussed in the above subsection. In order to obtain a consistent result for the two flux operators $\hat{\tilde{E}}^{\rm Alt}_\mu(S_t)$ and $\hat{\tilde{E}}^{\rm Fun}_\mu(S_t)$ acting on the same state $\left|{\left({}^{\sst (u/d)}\!\beta_t\right)^{J=0}}_{M=0}\right\rangle$ for any $t\in(0,1)$, corresponding to the situations that the intersecting point $\tilde{v}_t$ locates at any interior point of ${}^{\sst (u/d)}\!e$, the factor $\kappa_{\rm reg}$ should be fixed as $\frac12$ from the results in Eqs. \eqref{Alt-action-graph-up} and \eqref{Alt-action-graph-down}. It is easy to see that the consistent result will also be kept if there are more than one edges of $\gamma$ intersecting $S_t$ at their interior points. Hence the answer is affirmative for the first case.

The second case is that all edges of $\gamma$ belong to the type out with respect to a surface, and in this case there is no intersection point. It is easy to see that, in this case, the actions of two flux operators vanish, and thus are consistent.

\begin{figure}[t]
\centering
\begin{minipage}[l]{0.2\textwidth}
\centering
\includegraphics[width=1.4in]{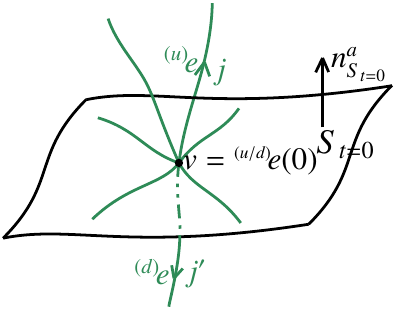}\\
(a)
\end{minipage}
\hspace{0.1cm}
\begin{minipage}[l]{0.2\textwidth}
\includegraphics[width=1.7in]{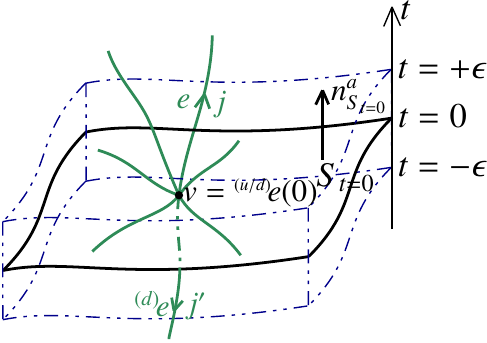}\\
(b)
\end{minipage}
\caption{(a) A surface $S_{t=0}$ intersects a graph $\gamma$ at the end points ${}^{\sst (u/d)}\!e(t=0)$ of edges ${}^{\sst (u)}\!e$ and ${}^{\sst (d)}\!e$ in $\gamma$. (b) The surface $S_{t=0}$ is modified as a region consists of a family of $\{S_t\}$ with $t\in (-\epsilon,+\epsilon)$ for a small enough parameter $\epsilon$.}
\label{modify}
\end{figure}

Now let us consider the third case in which the intersection points between $S_t$ and $\gamma$ locate at the end points, rather than interior points, of the edges of $\gamma$, as shown in Fig. \ref{modify}. Let us firstly analyze the difference between the two flux operator defined in Eqs. \eqref{Alt-flux-tau-def} and \eqref{Fun-flux-tau-def}, focusing on the ways of their action. Essentially, the two flux operators extract the information of quantum states by the right-invariant vector fields. In the first case, an original edge ${}^{\sst (u/d)}\!e$ was divided by $S_t$ at an interior intersection point $\tilde{v}_t$ into two edges ${}^{\sst (u/d)}\!e^t_1$ and ${}^{\sst (u/d)}\!e^t_2$ which are linearly dependent at  $\tilde{v}_t$. The linear dependence of ${}^{\sst (u/d)}\!e^t_1$ and ${}^{\sst (u/d)}\!e^t_2$ at $\tilde{v}_t$ ensures that only two terms, each term only involving an edge ${}^{\sst (u/d)}\!e^t_I$ for the given graph and two new additional edges, contribute to the sum of $\hat{Q}_{\tilde{v}_t}$ in Eq. \eqref{Q-sum} for the reduced expression \eqref{Alt-tau-sim-action} of \eqref{Alt-flux-tau-def}. In other words, the action of $\hat{\tilde{E}}^{\rm Alt}_\mu(S_t)$ consists of two terms, in which each term only contains the information of ${}^{\sst (u/d)}\!e^t_1$ or ${}^{\sst (u/d)}\!e^t_2$. On the other hand, it is easy to see that $\hat{\tilde{E}}^{\rm Alt}_\mu(S_t)$ has the same way of action from Eq. \eqref{Fun-action-simp}. Hence it is not surprising to us that the two flux operators are also consistent to each other for this case. However, the situation in the third case differs from that in the first case, because in the former the intersection points are the end points, at which other edges of $\gamma$ may also intersect. Thus the sum in the volume operator appeared in Eqs. \eqref{Alt-flux-tau-def} should take over all possible triplets of edges with the intersection point as their end points, rather than takes over the triplets in which each triplet only consists of an edge of $\gamma$ and two new additional edges. In other words, the action of $\hat{\tilde{E}}^{\rm Alt}_\mu(S_t)$ in Eq. \eqref{Alt-flux-tau-def} will mix the informations of different edges of $\gamma$ by the action of the volume operator. Apparently, the coupled action of the alternative flux operator in Eq. \eqref{Alt-flux-tau-def} on edges differs from the linear decoupled action of the fundamental flux operator in Eq. \eqref{Fun-flux-tau-def}. Hence in order to obtain a consistent action of the two flux operators, one has to slightly modify the definition of the alternative flux operator in Eq. \eqref{Alt-flux-tau-def} in a way that its actions on different edges are decoupled. A strategy was proposed in \cite{Giesel:2005bk,Giesel:2005bm} by redefining the alternative flux operator as a limiting operator, which is discussed as follows.

Without loss of generality, we consider that a general graph $\gamma$ intersects a surface $S_{t=0}$ at a vertex $v$, shown in the Fig. \ref{modify}. The corresponding spin network function is denoted by 
\begin{align}
 \left|T_{\gamma}^{\left(v,{}^{\sst (u)}\!e,{}^{\sst (d)}\!e\right)}\right\rangle&=\makeSymbol{
\includegraphics[height=6cm]{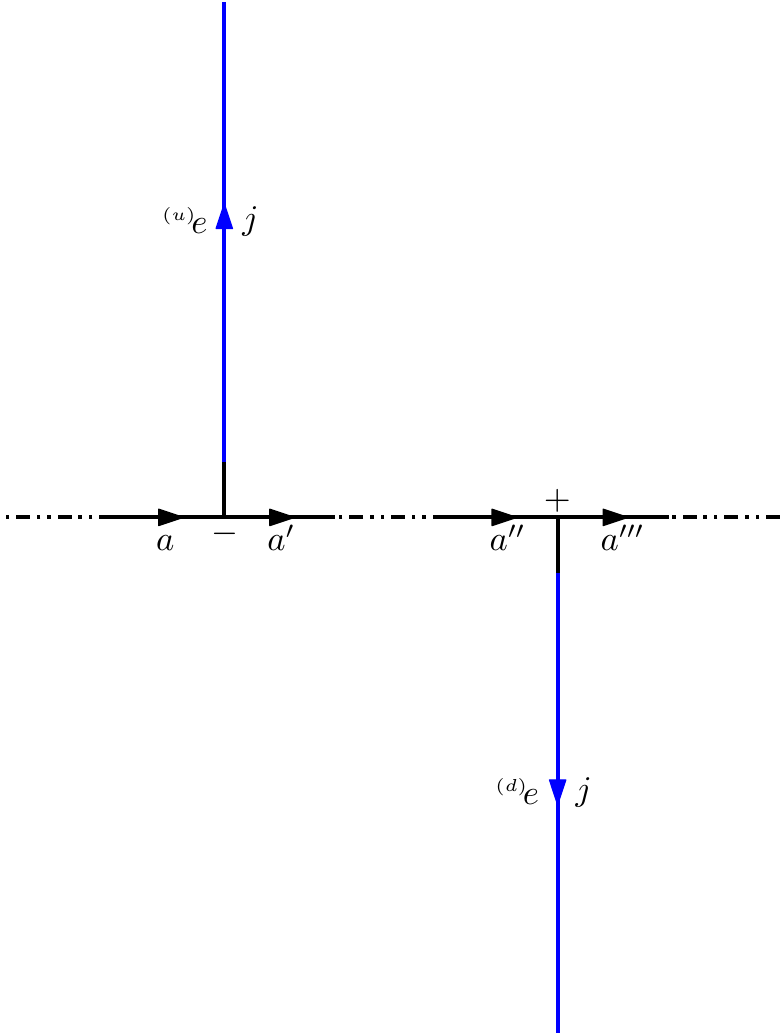}}\,.
\end{align}

By modifying the surface $S_{t=0}$ as a region consists of a family of $\{S_t\}$ with $t\in (-\epsilon,+\epsilon)$ for a small enough parameter $\epsilon$ such that there are no more vertices of $\gamma$ contained in this region, the alternative flux operator is defined as the following limitation \cite{Giesel:2005bk,Giesel:2005bm}
\begin{align}\label{limit-flux-def}
{}^{({\rm lim})}\!\hat{\tilde{E}}^{\rm Alt}_{\mu}(S_{t=0})\,\left|T_{\gamma}^{\left(v,{}^{\sst (u)}\!e,{}^{\sst (d)}\!e\right)}\right\rangle&:=\lim_{\epsilon\rightarrow0}\frac{1}{2\epsilon}\int_{-\epsilon}^{+\epsilon}{\rm d}t\hat{\tilde{E}}^{\rm Alt}_\mu(S_t)\,\left|T_{\gamma}^{\left(v,{}^{\sst (u)}\!e,{}^{\sst (d)}\!e\right)}\right\rangle\notag\\
&=\lim_{\epsilon\rightarrow0}\frac{1}{2\epsilon}\left[\int_{0}^{+\epsilon}{\rm d}t\hat{\tilde{E}}^{\rm Alt}_\mu(S_t)\,\left|T_{\gamma}^{\left(v,{}^{\sst (u)}\!e,{}^{\sst (d)}\!e\right)}\right\rangle+\int_{-\epsilon}^{0}{\rm d}t\hat{\tilde{E}}^{\rm Alt}_\mu(S_t)\,\left|T_{\gamma}^{\left(v,{}^{\sst (u)}\!e,{}^{\sst (d)}\!e\right)}\right\rangle\right]\,,
\end{align}
where ${\rm d}t$ denotes the Lebesgue measure, $\hat{\tilde{E}}^{\rm Alt}_\mu(S_t)$ is defined in Eq. \eqref{Alt-flux-tau-def}. The advantages of the limitation operator are in twofolds. First, the actions of the flux operator at the intersection points of the edges with the surface $S_t$ for any $t\neq0$ are decoupled with each other. Second, the contribution to the actions of the flux operator at the surface $S_t$ with $t=0$ corresponds to a measure zero set in the integral and hence can be removed from the actions. The action \eqref{limit-flux-def} can be calculated by graphical method as
\begin{widetext}
\begin{align}
&{}^{({\rm lim})}\!\hat{\tilde{E}}^{\rm Alt}_{\mu}(S_{t=0})\,\left|T_{\gamma}^{\left(v,{}^{\sst (u)}\!e,{}^{\sst (d)}\!e\right)}\right\rangle\notag\\
&=\lim_{\epsilon\rightarrow0}\frac{-2\kappa_{\rm reg}\ell_{\rm p}^2\,\beta\,\chi(j)}{2\epsilon}\left[\int_{0}^{+\epsilon}{\rm d}t\,\left(\makeSymbol{
\includegraphics[height=6cm]{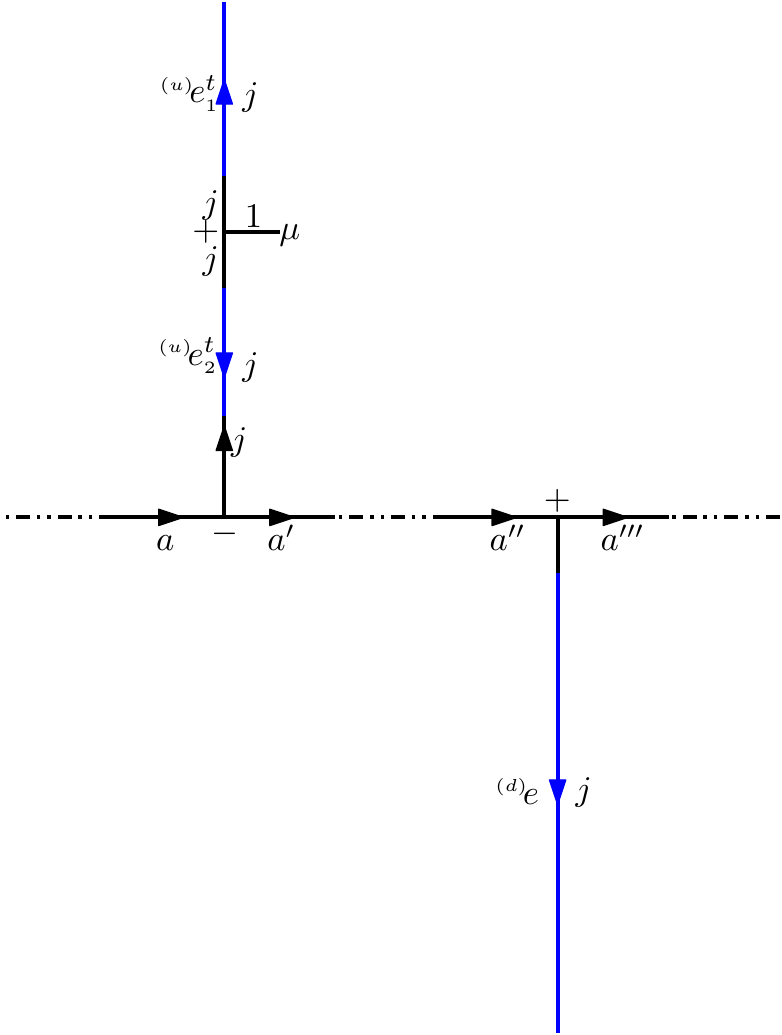}}\quad+\quad\cdots\right)+\left(\int_{-\epsilon}^{0}{\rm d}t\,\makeSymbol{
\includegraphics[height=6cm]{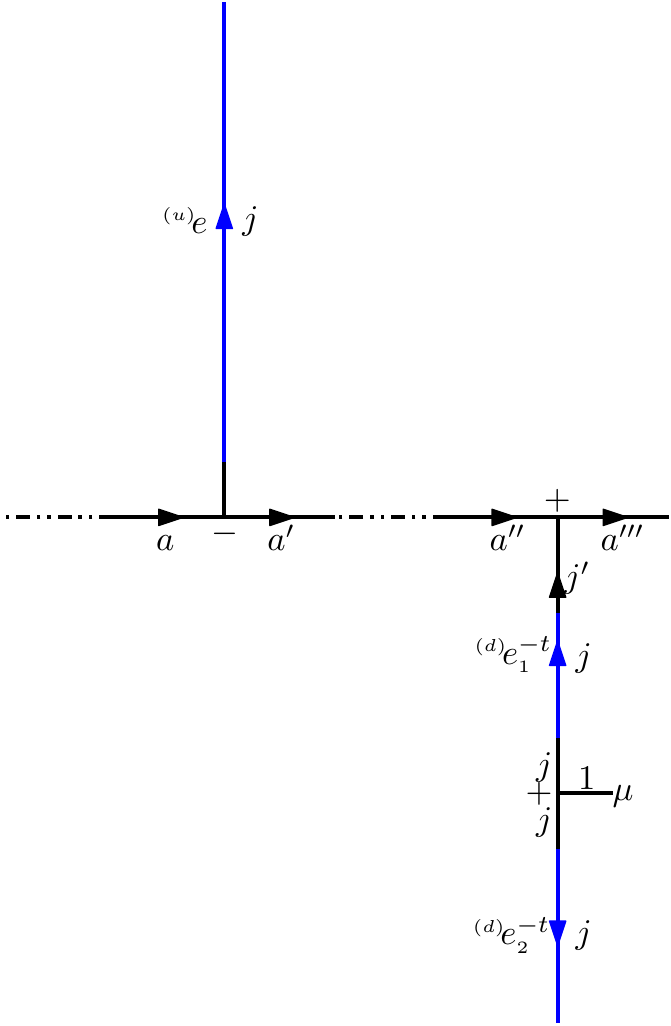}}\quad+\quad\cdots\right)\right]\notag\\
&=\frac{-2\kappa_{\rm reg}\ell_{\rm p}^2\,\beta\,\chi(j)}{2}\lim_{\epsilon\rightarrow0}\left[\,\left(\makeSymbol{
\includegraphics[height=6cm]{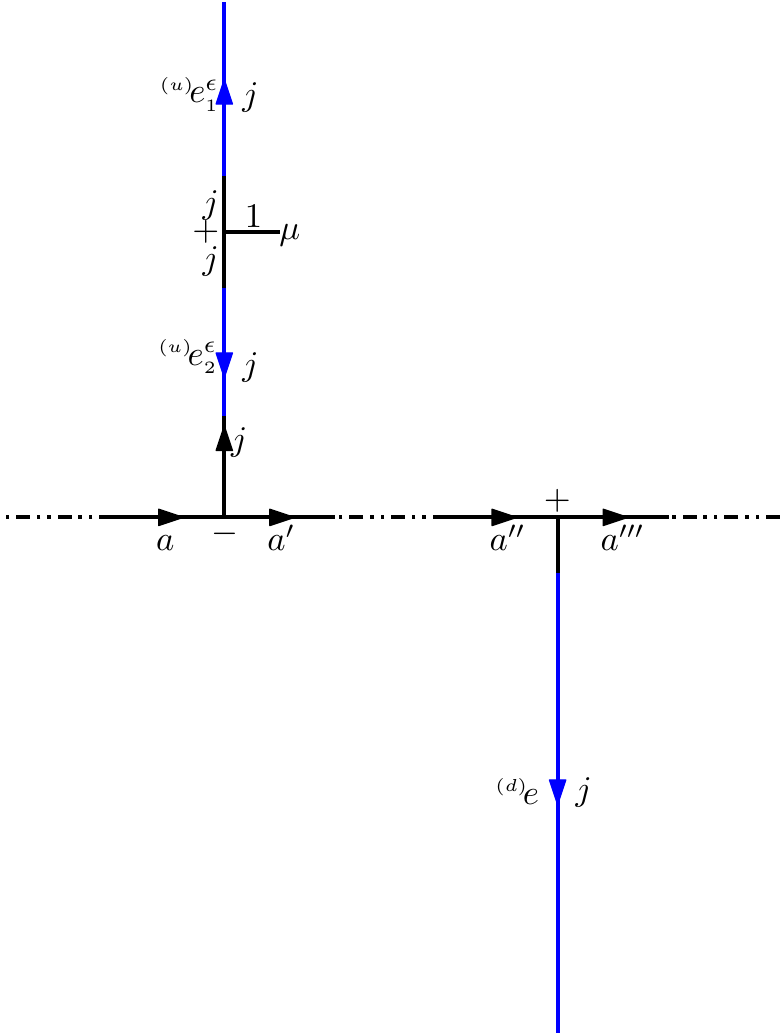}}\quad+\quad\cdots\right)+\left(\,\makeSymbol{
\includegraphics[height=6cm]{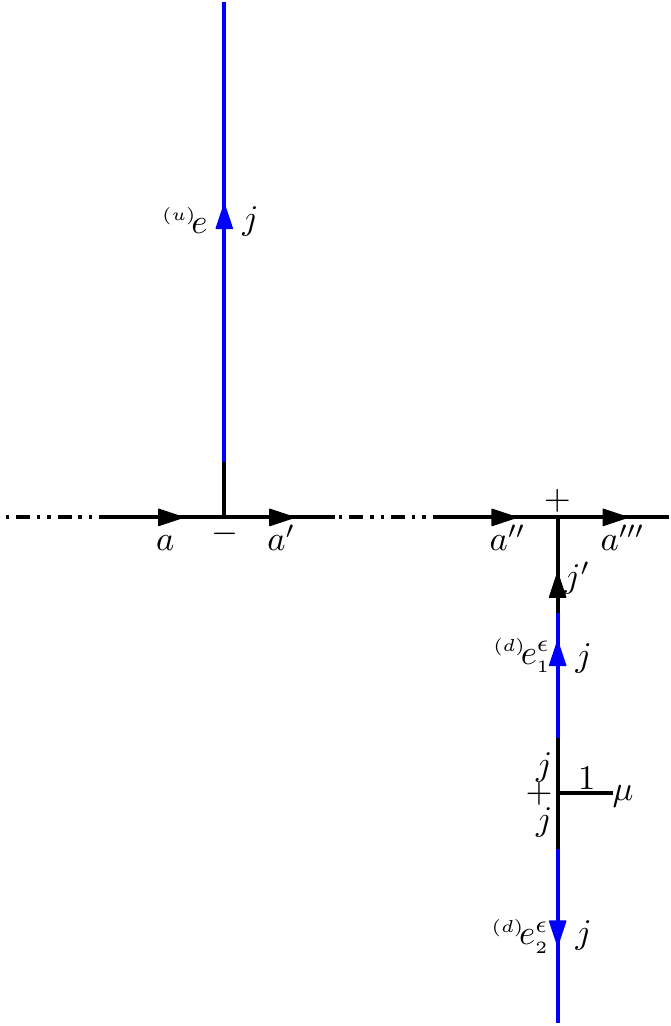}}\quad+\quad\cdots\right)\right]\notag\\
&=\frac{-2\kappa_{\rm reg}\ell_{\rm p}^2\,\beta\,\chi(j)}{2}\left[\,\left(\makeSymbol{
\includegraphics[height=6cm]{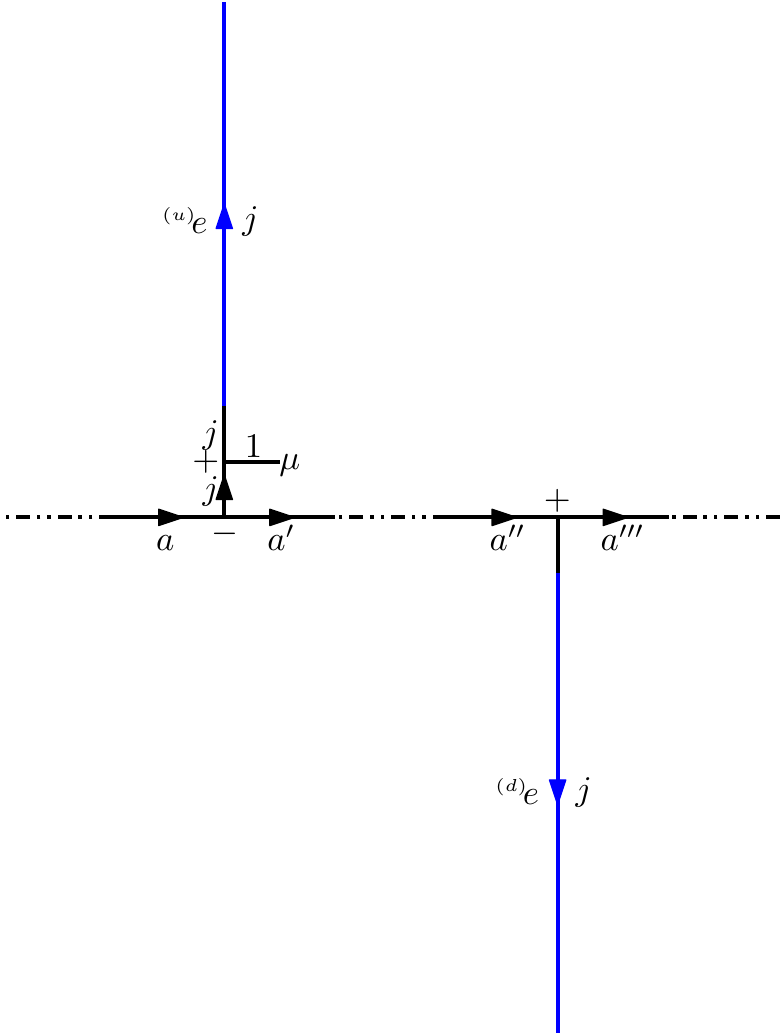}}\quad+\quad\cdots\right)+\left(\,\makeSymbol{
\includegraphics[height=6cm]{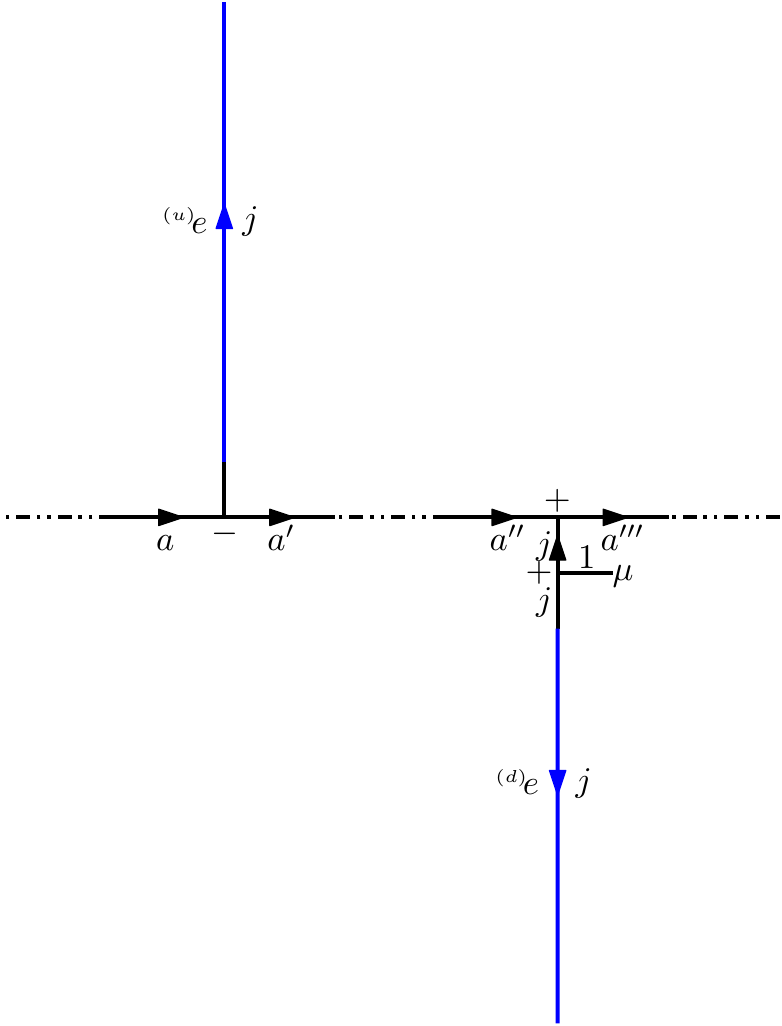}}\quad+\quad\cdots\right)\right]\notag\\
&=\frac{2\kappa_{\rm reg}\ell_{\rm p}^2\,\beta\,\chi(j)}{2}\left[\,\left(\makeSymbol{
\includegraphics[height=6cm]{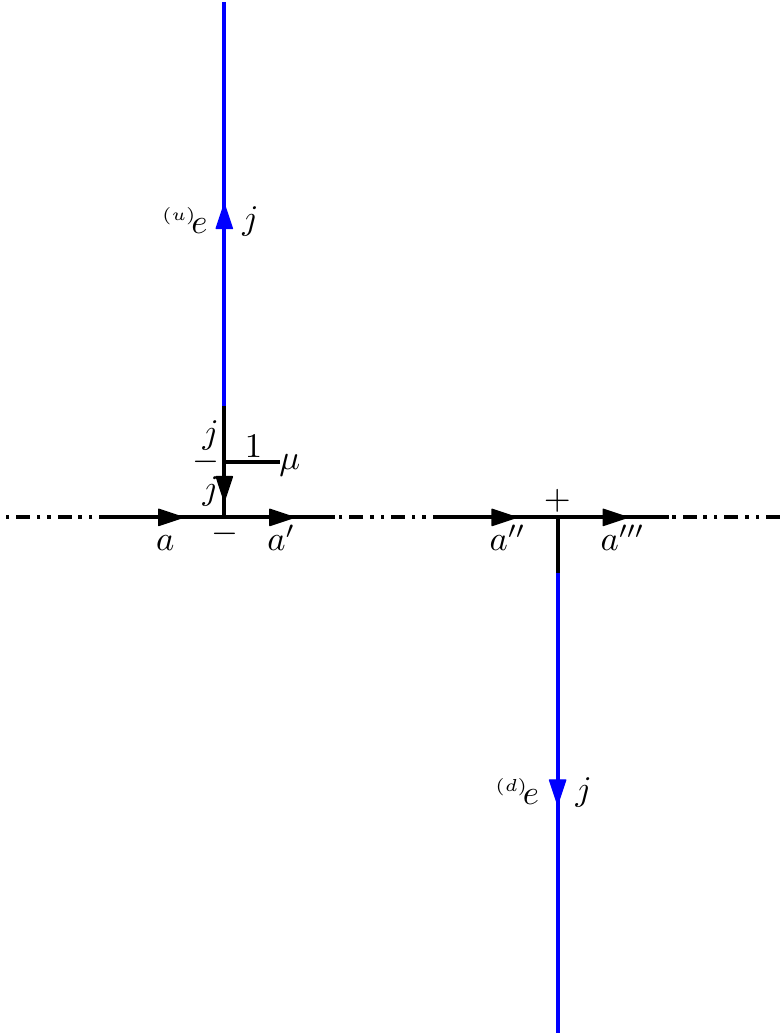}}\quad+\quad\cdots\right)+\left(-\,\makeSymbol{
\includegraphics[height=6cm]{figures/SNF-general-7}}\quad-\quad\cdots\right)\right]\notag\\
&=2\kappa_{\rm reg}\frac{\ell_{\rm p}^2\,\beta}{2}\left\{\left[J^\mu_{{}^{\sst (u)}\!e}\,\left|T_{\gamma}^{\left(v,{}^{\sst (u)}\!e,{}^{\sst (d)}\!e\right)}\right\rangle+\cdots\right]+\left[-J^\mu_{{}^{\sst (d)}\!e}\,\left|T_{\gamma}^{\left(v,{}^{\sst (u)}\!e,{}^{\sst (d)}\!e\right)}\right\rangle-\cdots\right]\right\}\notag\\
&=2\kappa_{\rm reg}\hat{\tilde{E}}^{\rm Fun}_{\mu}(S_{t=0})\,\left|T_{\gamma}^{\left(v,{}^{\sst (u)}\!e,{}^{\sst (d)}\!e\right)}\right\rangle\,,
\end{align}
\end{widetext}
where in the first step we have used Eqs. \eqref{Alt-action-graph-up} and \eqref{Alt-action-graph-down}. The above result shows that, for general case, the alternative flux operator ${}^{({\rm lim})}\!\hat{\tilde{E}}^{\rm Alt}_{\mu}(S)$ in Eq. \eqref{limit-flux-def} as a limiting operator is also consistent with the fundamental flux operator $\hat{\tilde{E}}^{\rm Fun}_{\mu}(S)$ in Eq. \eqref{Fun-flux-tau-def} if the factor $\kappa_{\rm reg}$ is fixed as $\frac12$.

\section{Summary and discussion}
It is well known that the triad operator plays an important role in the construction of Thiemann's Hamiltonian constraint operator. To test this quantization technique, an alternative flux operator was firstly constructed using the triad operator at the kinematical level in \cite{Giesel:2005bk,Giesel:2005bm}, and a consistency check on the fundamental and the alternative flux operators was also implemented. In this paper, we first introduced the construction of the fundamental and alternative flux operators, and then did the consistency check on them by employing the graphical calculus based on the original Brink graphical method. In order to obtain a consistent result for the actions of these operators on the same state, the following choices for the alternative flux operator were made: (i) the volume operator in Eq. \eqref{volume-operator} defined by Ashtekar and Lewandowski was chosen to construct the cotriad operator appearing in the alternative flux operator; (ii) a special operator ordering was used in Eq. \eqref{Alt-flux-tau-def}; (iii) the alternative flux operator was defined as a limitation shown in Eq. \eqref{limit-flux-def}.

By employing the consistent check in the graphical calculus, we fixed the factor $\kappa_{\rm reg}$ as $\frac12$, which differs from the one obtained in the algebraic calculation in \cite{Giesel:2005bk,Giesel:2005bm}. The relation between the factor $\kappa_{\rm reg}$ in this paper and the factor $C_{\rm reg}$ in \cite{Giesel:2005bk,Giesel:2005bm} is $\kappa_{\rm reg}=48 C_{\rm reg}$. This difference comes from the different regularizations for the alternative flux operator used in the present paper and Ref. \cite{Giesel:2005bm}. In the regularization used in Ref. \cite{Giesel:2005bm}, the action of the alternative flux operator on a state with an edge puncturing the two-surface $S$ of the flux equals to its action on a state with an up-type edge with respect to $S$ \cite{Giesel:2005bk}. This treatment is not consistent with that of the standard flux operator. Had a consistent treatment was taken for the two flux operators, the algebraic calculation would give the same results of $\kappa_{\rm reg}=\frac12$ as ours. In this sense, our calculation also confirms the consistency of the graphical method and the algebraic method in LQG, in addition to the consistency of the fundamental and alternative flux operators.

\begin{acknowledgments}
This work is supported in part by NSFC Grants No. 11765006, No. 11875006, and No. 11961131013.
\end{acknowledgments}

%========================================================================================================================

%
%========================================================================================================================

\end{document}